\documentclass[a4paper,fleqn,usenatbib]{mnras}

\usepackage[T1]{fontenc}
\usepackage{ae,aecompl}

\usepackage{graphicx}

\usepackage{amsmath}	
\usepackage{amssymb}	
\usepackage{hyperref}
\usepackage{fixltx2e}   

\newcommand{\sigmath}{\ensuremath \sigma_{\!*}}
\newcommand{\sig}{$\sigmath$}

\newcommand{\gaussfs}{$\hat f\!\sigmath$}
\newcommand{\hermitefs}{$\tilde f\!\sigmath$}

\newcommand{\gadget}{GADGET-3}
\newcommand{\GSnap}{\textsc{GSnap}}

\newcommand{\masssig}{$m\sigmath$}
\newcommand{\fluxsig}{$f\!\sigmath$}
\newcommand{\sunrise}{\textsc{Sunrise}}

\newcommand{\eg}{e.g.,}
\newcommand{\ie}{i.e.,}

\newcommand{\widefigure}[2] {\centering \leavevmode \includegraphics[width={#2\textwidth}]{#1}}
\newcommand{\narrowfigure}[2] {\centering \leavevmode \includegraphics[width={#2\columnwidth}]{#1}}

\title[The Effects of Dust and Star Formation on Apparent Velocity Dispersion]{Stellar Velocity 
Dispersion in Mergers: The Effects of Dust and Star Formation}

\author[N.~R.~Stickley \& G.~Canalizo]{Nathaniel~R.~Stickley\thanks{E-mail: nstic001@ucr.edu}
and
Gabriela~Canalizo\thanks{E-mail: gabriela.canalizo@ucr.edu}\\
Department of Physics and Astronomy, University of California, 900 University Ave, Riverside,
CA 92521 USA}

\begin{document}



\maketitle

\begin{abstract}
We investigate the effects of stellar evolution and dust on measurements of stellar
velocity dispersion in mergers of disk galaxies. \textit{N}-body simulations and
radiative transfer analysis software are used to obtain mass-weighted and flux-weighted
measurements of stellar velocity dispersion. We find that the distribution of dust with
respect to the distribution of young stars in such systems is more important than the
total degree of attenuation. The presence of dust typically causes flux-weighted
measurements of stellar velocity dispersion to be elevated with respect to mass-weighted
measurements because dust preferentially obscures young stars, which tend to be
dynamically cooler than older stellar populations in such systems. In exceptional
situations, in which young stars are not preferentially obscured by dust, flux-weighted
velocity dispersion measurements tend to be negatively offset with respect to
mass-weighted measurements because the dynamically cool young stellar populations are more
luminous, per unit mass, than older stellar populations. Our findings provide a context
for comparing observationally-obtained measurements of velocity dispersion with
measurements of velocity dispersion obtained from galaxy merger simulations.
\end{abstract}


\section{INTRODUCTION}

Central stellar velocity dispersion (\sig) is an important observable quantity in
galactic astronomy because it correlates with many other galactic properties, such as
depth of the galaxy's potential well, the mass of its central supermassive black
hole \citep{ferrarese2000, gebhardt2000, tremaine2002}, its luminosity, and its surface
brightness distribution \citep{davies1987, dressler1987, bender1992}.

Although \sig\ has been measured from observations and in numerical simulations for
decades, little work has been done to compare the fundamentally different methods
that are used to measure \sig\ in real galaxies versus simulated galaxies. Thus,
comparing values of \sig\ in the galaxy simulation literature with
observationally-obtained measurements of \sig\ has been problematic.

Observationally, \sig\ can only be measured by analysing the light emitted by a galaxy.
This is an inherently flux-weighted measurement method. In contrast, the natural way of
measuring \sig\ in a numerical simulation involves using the velocity and mass data in the
simulation to compute a mass-weighted velocity dispersion. This mass-weighted measurement
technique has been used for as long as \sig\ has been measured
in simulations\citep[\eg][]{villumsen1982, cox2006a, robertson2006a, johansson2009}. If
real galaxies consisted of stars with a uniform mass-to-light ratio ($\Upsilon$) and
contained no gas nor dust, observations of \sig\ should, in principle, agree perfectly
with the quantity computed from simulation data. However, reality is not so simple;
galaxies typically contain a broad variety of stars and at least a small amount of dust.

In our earlier work \citep[][hereafter denoted SC1]{stickley2012}, we found that including
a toy model for dust attenuation---a uniformly dense slab of attenuating material---could
potentially cause a decrease in the flux-weighted velocity dispersion (\fluxsig), relative
to the mass-weighted velocity dispersion (\masssig) in a simulated galaxy. Similar
results were obtained analytically by \citet{baes2000}, who studied the effect of
diffusely distributed dust on measurements of stellar kinematics in elliptical galaxies.
Unlike our toy model and the smooth distributions of dust assumed by \citet{baes2000},
real interstellar dust is not smoothly distributed; it occurs in clumps, sheets, and
filaments as well as in a more diffuse state with varying density. Thus, based on previous
work, it is not clear how a realistic dust distribution would affect the value of
\fluxsig.

Even in the absence of dust, comparisons of \fluxsig\ and \masssig\ are not
straightforward because stars in galaxies are not uniformly luminous. In a previous paper
\citep[][hereafter denoted SC2]{stickley2014}, we found that measurements of \masssig\
that were based upon newly-formed stars in simulated galaxies were significantly lower
than measurements of \masssig\ that included stars of all ages. This finding was
consistent with the observations of the so-called ``\sig\ discrepancy''
\citep{rothberg2010,rothberg2013}, in which different stellar populations in a single
galaxy yield discrepant values if \sig. Since young stellar populations are more luminous
per unit mass (i.e., have smaller $\Upsilon$), the presence of young stars may weight the
value of \fluxsig\ downward relative to \masssig.

In this paper, we examine the effects of dust and stellar evolution on measurements of
\fluxsig\ in order to gain insights into the differences between the value of \sig\
measured observationally and the value of \sig\ reported in the galaxy simulation
literature. We describe our methods in Section~\ref{section:methodology} and present our
findings in Section~\ref{section:results}. General conclusions and limitations to the
work are discussed in Section~\ref{section:discussion}.

\section{METHODOLOGY}\label{section:methodology}

We began by selecting a small set of snapshots from a binary galaxy merger simulation that was 
previously analysed in SC2. Using these snapshots, we generated synthetic Doppler-broadened galaxy 
spectra. We analysed these synthetic spectra to obtain flux-weighted velocity dispersion 
measurements. Then, the particle mass and velocity information in the snapshots was analysed to 
obtain mass-weighted velocity dispersion measurements. A method of consistently comparing the 
mass-based and flux-based measurement methods was developed so that intrinsic, systematic effects 
could be separated from the effects of dust and star formation.

\subsection{Numerical Simulations}

We have used simulation snapshots from merger S1, described in SC2. Since the details of
this simulation have already been discuss in detail, we will only summarize the key
features here; refer to SC2 for the full details of the simulation.

The simulation was a binary, 1:1 mass ratio, prograde-prograde merger of disk galaxies. It
was performed using the $N$-body, SPH code, \gadget\ \citep{gadget}. Each progenitor
consisted of $1.6\times10^6$ particles. Eighty percent of the disk mass of each progenitor
galaxy initially consisted of stars with the other 20\% in the form of gas and dust. The
gravitational softening length of stars was 25~pc, meaning that physical processes on
scales smaller than 25~pc could not be resolved. Each stellar particle in the simulation
represented a population of stars, rather than individual stars. The interstellar medium
was modelled as a multi-phase gas with hot and cool phases. The hot gas was able to cool
radiatively, while cool gas was able become heated by stellar feedback and AGN feedback. A 
sub-resolution approximation was used to include stellar formation in
the simulation; new stellar particles were spawned in cool, dense regions of gas at a rate
designed to match observational evidence. As stars evolved and added metals to the ISM,
the metallicity of the gas increased. Each stellar particle that formed during the
simulation carried a variable specifying its creation time and metallicity. The latter was
set equal to the metallicity of the gas from which the particle was spawned.

\subsection{Mass-Weighted Velocity Dispersion}

The mass-weighted velocity dispersion (\masssig) was computed with
\GSnap\footnote{\url{http://www.gsnap.org}} (N.R. Stickley, in preparation), using the
same
technique described in SC1 and SC2. In summary, a virtual rectangular slit of width $w = 2$~kpc and length $\ell =
20$~kpc was placed on the galaxy of interest. A viewing direction, ($\theta, \phi$) and slit position angle $\alpha$,
were then specified. The masses and velocities of all stars appearing in the slit were used to compute \masssig,
according to

\begin{equation}
m\sigmath = \sqrt{v_i^2m_i/M - (v_im_i/M)^2}
\label{eq:ms}
\end{equation}
\noindent
with
\begin{displaymath}
 M=\sum_i m_i
\end{displaymath}
\noindent
where the standard summation convention has been utilized; repeated indices imply a sum over that 
index. Using this technique allowed us to directly compare the present work with the results 
presented in SC2. The primary source of uncertainty in this measurement was particle noise, which 
never exceeded 0.9\% of the measured value. Note that this measurement technique can confuse 
velocity dispersion with rotation.

\subsection{Flux-Weighted Velocity Dispersion}

In this section, we describe the processes that were used to obtain and analyse the
synthetic, Doppler-broadened, galaxy spectra.

\subsubsection{Simulated Spectra}

Synthetic, Doppler-broadened galaxy spectra were generated using the polychromatic, Monte Carlo, 
radiative transfer code, \sunrise\ \citep{jonsson2006, jonsson2010a, jonsson2010b}.  \sunrise\ was 
capable of creating realistic images of galaxies from arbitrary viewing directions. More 
importantly, the code could compute a high resolution spectrum for each pixel of each image that it 
generated.

Given a \gadget\ snapshot file, \sunrise\ began by discretising the spatial domain of the simulation 
using an adaptive mesh. A Monte Carlo radiative transfer algorithm was then performed on the 
discretised volume. The algorithm assumed that the dust content of the ISM was proportional to the 
metallicity of the SPH particles in the simulation snapshot file. In this work, we assumed that 40\% 
of the metals in the ISM occurred in the form of dust. Because the stellar particles in our galaxy 
simulation represented entire stellar populations, rather than individual stars, each stellar 
particle in \sunrise\ emitted a spectrum corresponding to a population of stars. These spectra were 
pre-computed using \textsc{Starburst99} \citep{leitherer1999}. Since our \gadget\ simulations only 
tracked the ages and metallicities of stellar particles that formed during the simulation, we 
manually assigned ages and metallicities to the pre-existing stellar particles; all particles were 
assumed to form instantaneously, 6~Gyr before the beginning of the simulation with a metallicity of 
$Z = 0.025$, which is approximately $1.25~Z_\odot$. \sunrise\ treated stellar particles younger than 
10~Myr as active star-forming regions, containing enhanced dust concentrations and a 
photodissociation region. These star-forming regions are modelled using the code, 
\textsc{MappingsIII} \citep{dopita2005,groves2008}. AGN emission was not included in the radiative 
transfer computation.

Using the stellar spectra and the spatial distribution of stars and dust, \sunrise\ computed the 
resulting spectra by tracing $10^7$ ``photon bundles'' through the simulation domain. Each photon 
bundle carried spectral data on a random walk through the simulated galaxy and accounted for Doppler 
shifts due to the motion of the individual stellar particles. The bundles were eventually collected 
by a virtual integral field spectrograph. As the photons were collected, the Doppler shifts of the 
individual bundles combined to form Doppler-broadened spectra. In addition to computing the 
dust-attenuated spectral information for each pixel of the generated image, \sunrise\ computed the 
unattenuated spectra. This allowed us to clearly identify the effect of attenuation in each virtual 
observation, by comparing attenuated and unattenuated fluxes.

Besides the galaxy simulation snapshot file and the input spectra, the other main user input to the 
code was the choice of dust model (i.e., the dust grain size distribution and albedo). For this 
work, we used the Milky Way dust model from \citet{weingartner2001}. It should be noted that the 
dust model used by \sunrise\ significantly affects the final spectrum of the simulated galaxy 
\citep{jonsson2010a}.

The radiative transfer computation required a large amount of memory, partially owing to the fact 
each photon bundle carried a detailed spectrum. In order to reduce the memory requirement and 
accelerate the computation, we limited our spectral coverage to the Mg~Ib region, from $5040$~\AA\ 
to $5430$~\AA. Our input stellar spectra contained 1170 wavelength bins in this region, 
logarithmically spaced. Once \sunrise\ had finished generating synthetic spectra for the individual 
pixels of each image, we placed a rectangular slit, measuring $2\times 20$~kpc, on the image. The 
spectra of all pixels appearing within the slit were then combined to form a single spectrum. This 
combined spectrum was analysed in order to determine the flux-weighted velocity dispersion 
(\fluxsig).

\subsubsection{Measuring Velocity Dispersion from Spectra \label{measuring-fluxsig}}

Stellar velocity information is encoded in all galaxy spectra, since the light emitted by a galaxy 
consists of the sum of the Doppler shifted spectra of its constituent stars. Similarly, each of our 
synthetic spectra consisted of a sum of Doppler-shifted particle spectra. In order to decode the 
spectra and recover the velocity information, we used the penalized pixel-fitting code, pPXF, 
developed by Cappellari \citep{cappellari2004, cappellari2012}. In general, the pPXF algorithm 
worked by fitting a parametrized model spectrum, $G_{\rm mod}(x)$, to an observed galaxy spectrum, 
$G(x)$. One of the parameters of $G_{\rm mod}(x)$ was \fluxsig. Thus, \fluxsig\ was ultimately 
determined by finding the model spectrum that best fit each galaxy spectrum. In our analysis, we 
made no attempt to correct for the effect of galaxy rotation.

More specifically, the model spectrum was a linear combination of template spectra convolved with a 
parametrized line of sight velocity distribution (LOSVD):

\begin{equation}
G_{\rm mod}(x) = \sum_{k=1}^K w_k [B*T_k](x) + \sum_{l=0}^L b_l\mathcal{P}_l(x)\qquad(w_k\geq0),
\end{equation}

\noindent
where $w_k$ are weights, $B(x) = \mathcal{L}(cx)$ is a broadening function, with $\mathcal{L}(v)$ 
the LOSVD, $c$ is the speed of light, $T_k$ is a library of template spectra, and $*$ denotes 
convolution. A linear combination of Legendre polynomials, $\mathcal{P}_l(x)$ (with weights $b_l$) 
was used to account for low-frequency differences between the shape of the templates and the shape 
of the galaxy spectrum.

The LOSVD was expanded as a Gauss-Hermite series,

\begin{equation}
\mathcal{L}(v) = \frac{\exp(-y^2/2)}{\sigma\sqrt{2\pi}}\left[ 1 + \sum_{m=3}^M h_m H_m(y)\right],
\end{equation}

\noindent
where $H_m$ are Hermite polynomials, $y \equiv (v-V)/\sigma$, and ($V, \sigma, h_3, h_4, \ldots, h_M 
$) are free parameters related to the moments of the velocity distribution. For example, $V$ is the 
mean line-of-sight velocity, $\sigma$ corresponds to the standard deviation (i.e., \fluxsig), $h_3$ 
is related to the skewness, and $h_4$ is related to the kurtosis. Optimal values for these 
parameters, as well as the weights, $w_k$ and $b_l$, were found using a non-linear least-squares 
minimization algorithm.

The quantity minimized by the least-squares optimization routine was the objective function:

\begin{equation}
\chi_{\rm p}^2 = \chi^2 (1 + \lambda^2\mathcal{D}^2), \label{objfunc}
\end{equation}

\noindent
where $\chi^2$ is given by

\begin{equation}
\chi^2 = \sum_{n=1}^N\left[\frac{G_{\rm mod}(x_n) - G(x_n)}{\Delta G(x_n)}\right]^2, \label{chisquared}
\end{equation}

\noindent
with $\Delta G(x_n)$ the measurement error on $G(x_n)$. The $\mathcal{D}^2$ in Eq.~(\ref{objfunc}) 
is a penalty term, given by the integrated square deviation of $\mathcal{L}(v)$ from its 
best-fitting Gaussian, $\mathcal{G}(v)$,

\begin{equation}
\mathcal{D}^2 = \frac{\int_{-\infty}^\infty [\mathcal{L}(v) - \mathcal{G}(v)]^2{\rm d}v}
                     {\int_{-\infty}^\infty\mathcal{G}^2(v){\rm d}v }
\end{equation}

\noindent
The penalty term clearly increases as $\mathcal{L}(v)$ deviates from a pure Gaussian. Thus, it has 
the effect of forcing the fitting routine to favour LOSVDs that are more nearly Gaussian. The 
parameter, $\lambda$, was a user-specified quantity that allowed us to adjust the importance of the 
penalty term. Setting $\lambda = 0$ caused the best-fitting LOSVD to be a general Gauss-Hermite 
series. Increasing the parameter $\lambda$ resulted in LOSVDs that were more nearly Gaussian. When 
using pPXF with noisy spectra, obtained observationally, non-zero values of $\lambda$ are often used 
in order to force the fitting routine to favour Gaussian LOSVDs. This helps pPXF to partially ignore 
the effect of noise in the spectra in cases for which the true LOSVD is Gaussian 
\citep{cappellari2004}. Our spectra did not include the sort of noise that is present in real (\ie\ 
observed) spectra, but we adjusted $\lambda$ nonetheless in order to examine the full range of 
velocity dispersions that could be obtained using our spectra. Specifically, we set $\lambda=0$ to 
determine the value of \fluxsig\ that would be measured by a researcher who prefers to approximate 
their LOSVDs using a Gauss-Hermite series. We also set $\lambda=10^6$ in order to determine the 
value of \fluxsig\ that would be reported by a researcher who prefers to use simple Gaussian LOSVDs.

We constructed a template library using a subset of the raw spectra emitted by the stellar particles 
in the \sunrise\ computation. The template library was gradually expanded to include more spectra 
until the effect of adding additional spectra no longer significantly effected the $\chi^2$ of the 
resulting fits. In total, 67 template spectra were included in the library. For the measurement 
error, $\Delta G(x_n)$, we assumed a uniform value for each wavelength bin in the spectrum. 
Specifically, we set each entry equal to 1\% of the mean signal strength (i.e., $\Delta G(x_n) = 
0.01\langle G\rangle$, for all $n$). This had no effect on the resulting values \fluxsig; it merely 
determined the magnitude of the formal measurement uncertainty.

\subsection{Consistent Comparisons}

In order to identify the effects of interstellar dust and star formation on measurements
of \fluxsig, we first needed a general method for consistently comparing measurements of
\masssig\ with measurements of \fluxsig. For instance, both measurement techniques needed
to include exactly the same region of space, measurement uncertainties needed to be
quantified, and intrinsic differences between the two measurement methods needed to be
identified.

\subsubsection{Slit Calibration and Antialiasing}

A virtual observatory was positioned 5~Mpc from the centre of the simulated system. Each observation 
was performed using a virtual integral field spectrograph with a resolution of $400\times400$ 
pixels. The field of view of each array was 200~kpc, thus each pixel represented a 
$0.5\times0.5$~kpc region of the simulation. In order to compare the mass-weighted and flux-weighted 
velocity dispersions, we first verified that we were measuring the same region of the simulation 
snapshot using our two methods. The verification process involved comparing detailed images, 
generated by \sunrise, with images generated using \GSnap's interactive particle visualization 
feature to confirm that the same particles were included in both types of measurements.

\begin{figure}
\centering
    \includegraphics[width=2in]{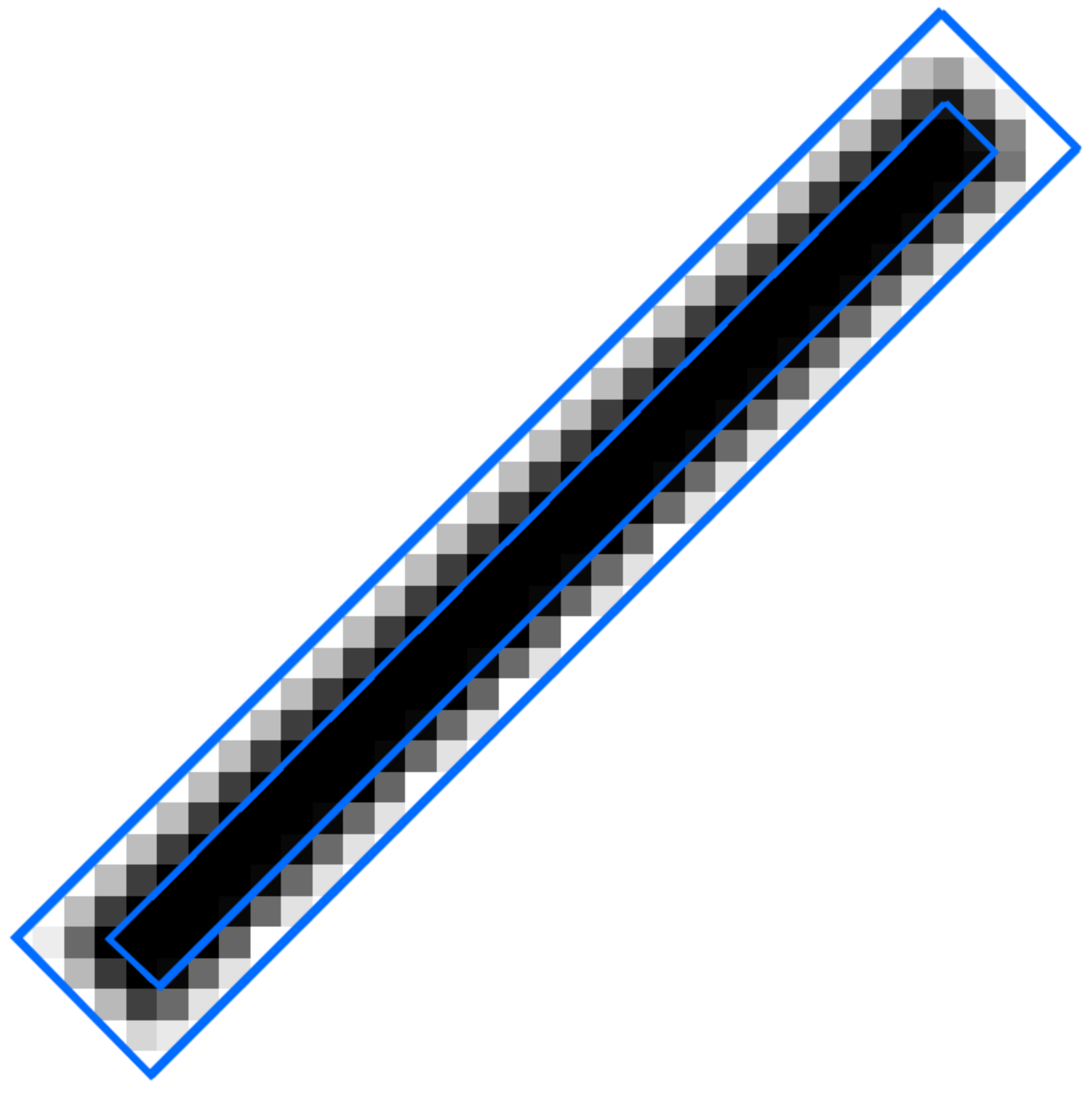}
\caption[The pixel weighting scheme]{\label{aliasing}An illustration of the
pixel-weighting scheme. White pixels indicate weights of zero while black pixels
indicate weights equal to unity. Gray pixels on the edge of the slit are weighted intermediately. The blue rectangles
indicate the smallest and largest slit sizes that were used when measuring \masssig. The fiducial slit (not shown) lies
mid-way between the blue rectangles.}
\end{figure}

Once the location and orientation of the slits had been calibrated, there remained a mismatch 
between the sizes of the flux slit and the mass slit, due to the pixelated nature of the flux data. 
Pixels falling on the border of the slit typically extended outside of the slit. Thus, flux from a 
large region outside of the slit contributed to the measurement of \fluxsig. To reduce this aliasing 
effect, we assigned weights to the pixels, depending on the degree of overlap with the slit. The 
result can be seen in Figure~\ref{aliasing}, where white corresponds to a weight of zero and black 
corresponds to a weight of unity. Edge pixels were weighted intermediately, so they appear as shades 
of grey. These weights were applied to the pixel spectra when the combined slit spectrum was 
computed. Furthermore, we varied the size of the mass slit in order to determine the maximum and 
minimum possible values of \masssig\ in the region that was measured by the flux-weighting method. 
The blue rectangles in Figure~\ref{aliasing} indicate the smallest and largest slit sizes that were 
used during this process. In Section~\ref{section:results}, we report the value of \masssig\ 
measured using the fiducial $2\times20$~kpc slit. We use the maximum and minimum values of \masssig\ 
over the full range of slit sizes as upper and lower bounds on the uncertainty whenever \masssig\ is 
compared with \fluxsig.

\subsubsection{An Intrinsic Measurement Discrepancy}

In order to characterize the intrinsic differences between the two measurement methods, we
compared measurements of \masssig\ and \fluxsig\ in snapshots consisting of identical
stars (i.e., stars of a specific mass, age, metallicity, and $\Upsilon$) and containing
no dust nor gas. The positions and velocities of the stars in these snapshots were
identical to the snapshots that were later analysed. In such systems, each star particle
contributed equally to the measurement of \masssig\ because the stellar masses were equal.
All stars emitted identical spectra and had identical luminosities. Thus, the inverse
square variation of flux with distance and the random noise introduced by the Monte Carlo
radiative transfer scheme used by \sunrise\ were the only effects that prevented each star
from contributing equally to the \fluxsig\ measurement (in the absence of random noise,
the stars nearest to the camera would have contributed approximately 4\%
more flux than the stars farthest from the camera).

Upon comparing the techniques, we found a systematic discrepancy. The flux-weighted
measurement technique yielded values that were elevated with respect to the fiducial
mass-weighted value by an average of 2.7\% when a Gauss-Hermite LOSVD was assumed and
3.3\% when a Gaussian LOSVD was assumed. After accounting for measurement uncertainty due
to slit mismatch, these offsets fell to 1.4\% and 2.0\%, respectively. The maximum offset
from the fiducial mass-weighted value was 5.8\% (Gauss-Hermite) and 6.2\% (Gaussian). No
negative offsets were observed; \textit{flux-based measurements always agreed with or
slightly exceeded the mass-based measurements}.

We do not believe that this discrepancy is a fundamental, real effect. It may have been
caused by an aspect of the spectral template library creation process. It also may have
been due to a peculiarity of the pPXF fitting code or the \sunrise\ radiative transfer
code. Note that, while it was not \textit{caused} by the pixel-weighting scheme, a
better pixel-weighting scheme may have been able to reduce the discrepancy. Regardless
of the exact cause, we were able to take the effect into account when performing our
later analysis because the effect was small and bounded in our set of observations. Any
measurement of \fluxsig\ that exceeded the corresponding \masssig\ measurement by more
than 6.2\% was attributable to the effects of dust attenuation or non-uniform $\Upsilon$.
Any negative offsets larger than the slit mismatch uncertainty could also be attributed to
these effects.

\subsection{Sample Selection \label{section:sample}}

In general, our observations were chosen to include a mixture of extreme and ordinary
situations. More precisely, snapshots that were known to exhibit extreme values of
\masssig\ were included along with snapshots that were known to be in a dynamically
relaxed state. Lines of sight known to have enhanced dust extinction (for instance,
parallel to a disk of gas) were included along with lines of sight with far less dust
extinction (\eg\ nearly perpendicular to disk structures). Intermediate cases were
included as well.

A total of seven snapshots from the simulation were examined using our mass-weighted and
flux-weighted velocity dispersion measurement methods. The first six of these were the
snapshots labelled $c$--$h$ in SC2. These correspond to interesting points in the dynamical
evolution of the merger. The seventh snapshot was the final snapshot of the simulation.
\sunrise\ was used to perform virtual observations on each snapshot along three viewing
directions. For each resulting image, we placed four slits---centred on the same pixel,
but rotated uniformly about that pixel's centre. Thus, we performed 21 virtual
observations and used 84 individual slits. The basic features of the snapshots are
summarized in Figure~\ref{summaryplot}. Detailed descriptions and renderings of each of
the 21 observations are presented in the \hyperref[section:appendix]{Appendix}. To gain a
better understanding of the detailed dynamics of the merger as a whole, refer to SC2.

\section{RESULTS}\label{section:results}

\subsection{Overview \label{section:overview}}

\begin{figure*}
\centering
    \widefigure{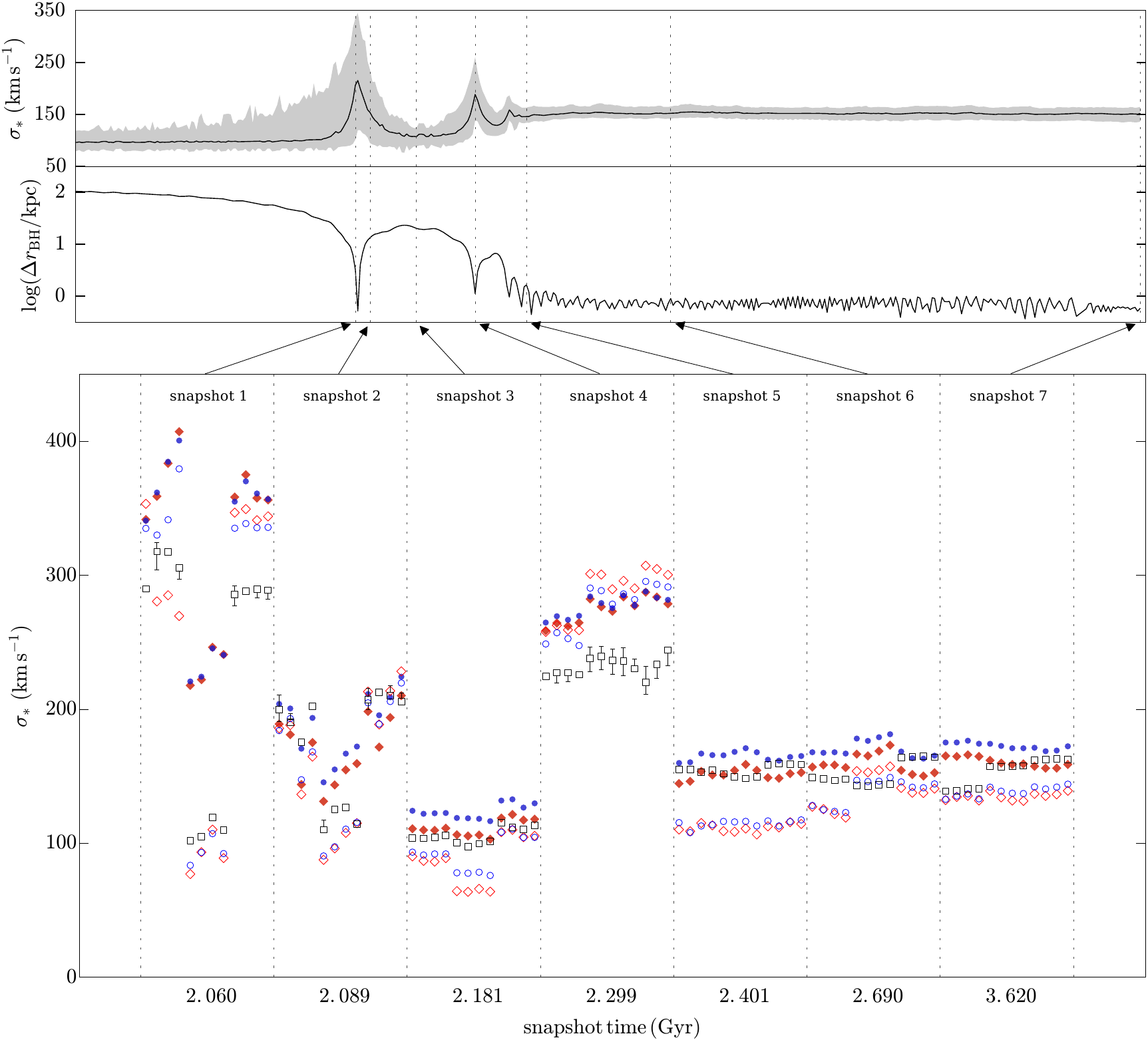}{0.9}
\caption[Overview of measurement results]{\label{summaryplot}Upper panel: the
mass-weighted \sig\ as a function of simulation time. The
dark line indicates the mean value of \sig, measured along 1000 random lines of sight. The upper and lower bounds of
the shaded region indicate the maximum and minimum values of \sig\ along the set of 1000 directions. Middle panel: the
separation distance between the supermassive black holes in the nuclei of the progenitor galaxies. Lower panel: An
overview of all velocity dispersion measurements performed on snapshots 1--7. Refer to Section~\ref{section:overview}
for the meanings of the symbols used in the plot.}
\end{figure*}

The lower panel of Figure~\ref{summaryplot} summarizes all of the velocity dispersion measurements 
performed on snapshots 1--7. Measurements of \masssig\ are shown as black boxes. Measurements of 
\fluxsig\ that assumed a pure Gaussian LOSVD, are plotted as blue circles. Measurements of \fluxsig\ 
that used the more general Gauss-Hermite series to model the LOSVD are plotted as red diamonds. 
Filled circles and diamonds indicate that the flux-weighted measurement was based on a spectrum that 
included dust attenuation. Open circles and diamonds indicate that dust was ignored when computing 
the spectrum. Error bars are only included in the plot when the measurement uncertainty is larger 
than the plotted symbol (this plotting convention is used throughout this paper). The horizontal 
axis indicates the simulation time at which the snapshot was saved. Within each snapshot bin, there 
are three groups of four measurements, corresponding to the three viewing directions and four slit 
orientations per viewing direction. Proceeding from left to right, the first four measurements were 
made along the direction of Camera~1, the second four represent Camera~2 measurements, and the final 
four represent Camera~3 measurements (refer to the \hyperref[section:appendix]{Appendix} for more 
information about the camera positions). For context, the top panel of Figure~\ref{summaryplot} 
shows the evolution of the mass-weighted \sig\ as a function of simulation time. The central panel 
indicates the separation distance between the supermassive black holes at the centres of the 
galaxies, which is a proxy for the nuclear separation distance.

In general, all measurement methods indicated that the velocity dispersion of the dynamically excited system in
Snapshot~4 was elevated with respect to the values in snapshots 3, 5, 6, and 7. It was
also clear that the systems in snapshots 5--7 were very similar to one another, which was
expected, since the merger remnant was passively evolving when the final three snapshots
were obtained.

Several trends are apparent in the data presented in Figure~\ref{summaryplot}. For instance, most flux-weighted
measurements that assumed a Gaussian LOSVD fell above the corresponding measurement that used a Gauss-Hermite series to
model the LOSVD. It was also clear that, in most cases, the \fluxsig\ measurements that included dust attenuation fell
above \masssig, while the non-attenuated \fluxsig\ typically fell below \masssig. These trends are examined in the
following sections.

\subsection{Gaussian versus Gauss-Hermite Fitting Methods}

Since researchers use different methods to extract the value of \fluxsig\ from real spectral data, we have examined the
differences between the two most common flux-based methods. As mentioned previously, these two methods differ due to the
assumed functional form of the LOSVD. On one extreme, a Gaussian LOSVD is assumed, on the other, a Gauss-Hermite series
is used to model the LOSVD. We will use the symbol, \gaussfs\  to denote a measurement made using the Gaussian method
and~\hermitefs\ to denote a measurement based on the Gauss-Hermite method.

\begin{figure}
\centering
    \narrowfigure{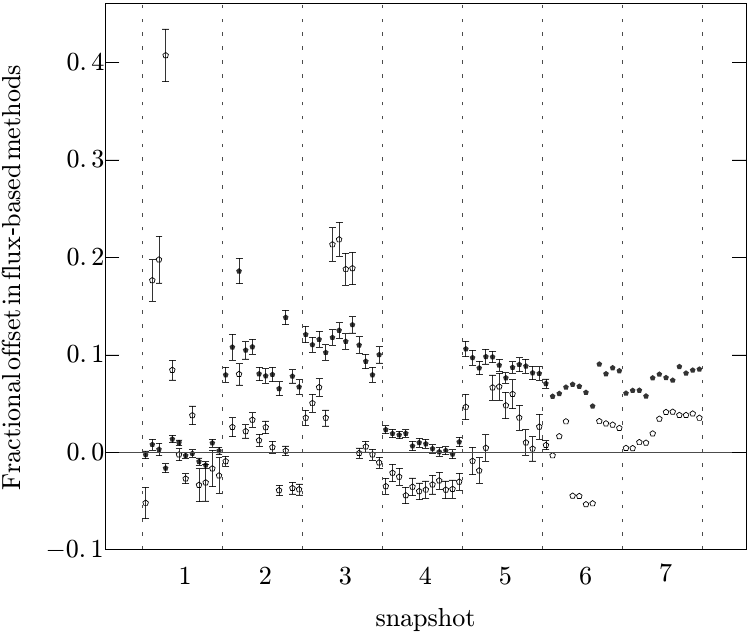}{1.0}
\caption[Fractional offset of Gaussian and Gauss-Hermite measurements]{The quantity  $(\hat f\!\sigmath-\tilde
f\!\sigmath)/\tilde f\!\sigmath$ for each measurement slit. This is the fractional offset
between flux-weighted velocity dispersions measured under the assumption of a Gaussian LOSVD (\gaussfs)
relative to those assuming a Gauss-Hermite LOSVD (\hermitefs). When dust attenuation was included in the
synthetic spectra (solid symbols), the Gaussian model consistently yielded larger
velocity dispersions in all snapshots except for Snapshot~1 and Snapshot~3. When dust
attenuation was not included (open symbols), there was no clear trend in the offset.}
\label{gauss-gauss-hermite}
\end{figure}

In Figure~\ref{gauss-gauss-hermite}, we plot the fractional offset between \gaussfs\ and \hermitefs. When dust
attenuation was not included in the spectra, there was no measurable, systematic offset
between the two methods. However, when dust attenuation was included, \gaussfs\ exceeded
 \hermitefs\ by 6.7\%, on average. This offset was not seen in some of the measurements
made on Snapshot~1 and Snapshot~3. Recall that these snapshots were
recorded during the first pass and between the first and second passes of the merger,
respectively.

\subsection{Mass-based versus Flux-based Measurements \label{section:ms_vs_fs}}

\begin{figure}
\centering
    \narrowfigure{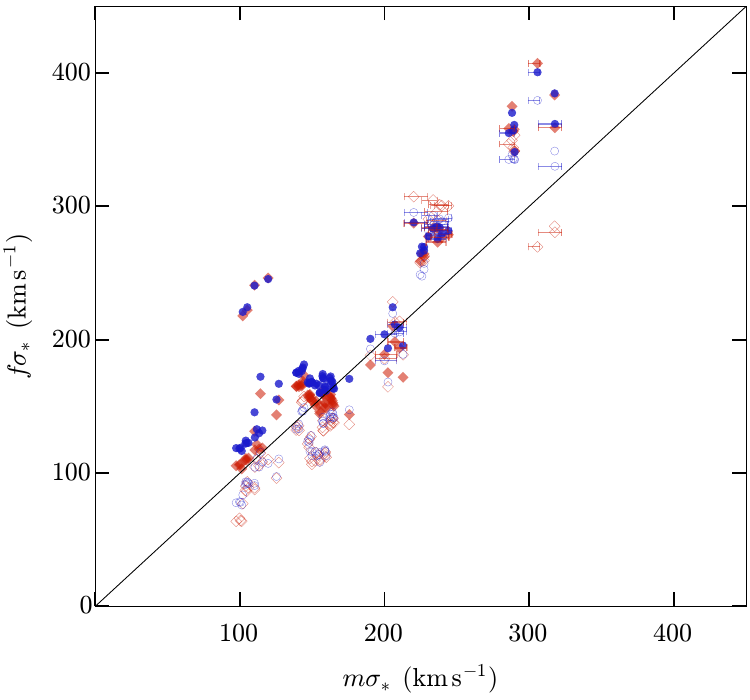}{1.0}
\caption[\fluxsig\ versus \masssig]{Flux-weighted versus mass-weighted velocity dispersion. The flux-weighted quantity
tends to be positively offset with respect to the mass-weighted quantity when dust is
included (solid circles and diamonds). The positive offset is accentuated when the LOSVD
is assumed to be Gaussian (blue circles). On the other hand, when dust
attenuation is not taken into account (open diamonds and circles), the flux-weighted quantity is negatively offset with
respect to the mass-weighted quantity.}
\label{fsvms}
\end{figure}

\begin{figure*}
\centering
    \widefigure{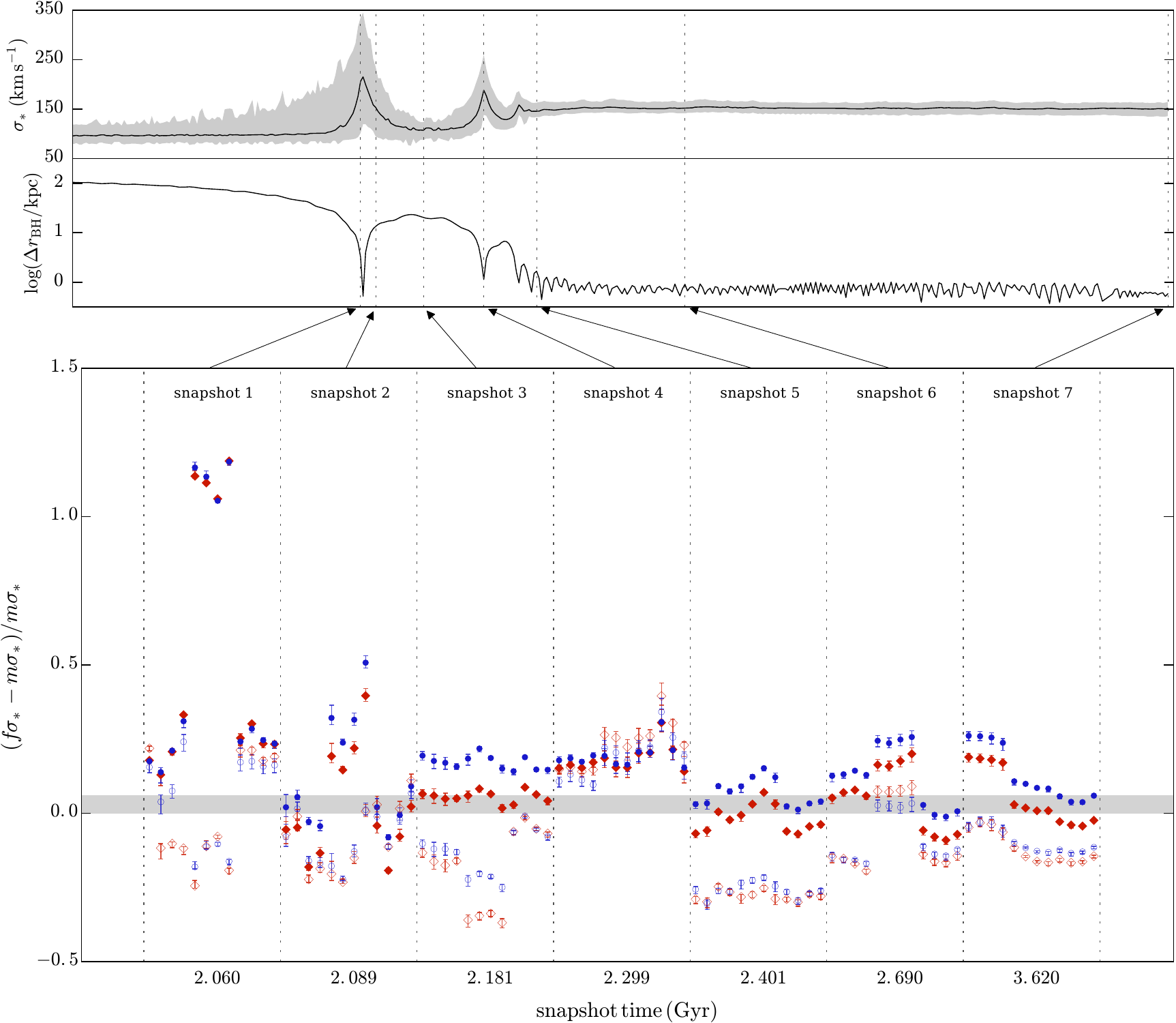}{0.9}
\caption{Upper panels: Refer to the caption of Figure~\ref{summaryplot}. Lower panel: The fractional offset between
flux-weighted and mass-weighted velocity dispersions. The shaded region indicates the range of offset values that may
have been caused by
the intrinsic discrepancy between the flux-weighted and mass-weighted measurement techniques. Only offsets falling
outside of this region were considered significant. When dust was included, a significant positive offset was observed
in a majority of the 84 measurement slits (63 \gaussfs\ measurements and 43 \hermitefs\
measurements). In the absence of dust, the offset was negative in the majority of
measurements (57 \gaussfs\ measurements and 58 \hermitefs\ measurements).}
\label{figure6}
\end{figure*}

In Figures~\ref{fsvms} and \ref{figure6}, we compare measurements of \masssig\ with
measurements of \fluxsig\ in the set of observations, described in
Section~\ref{section:sample}. For the measurements that included dust, the mean fractional
offsets between flux-weighted and mass-weighted velocity dispersions,
$(f\!\sigmath-m\sigmath)/m\sigmath$, were respectively 0.19 and 0.13 for \gaussfs\ and
\hermitefs. The offsets were largest along the direction of Camera~2 in Snapshot~1.
Interestingly, this particular camera direction was chosen, in part, because of the
apparently small amount of intervening dust along its line of sight. The large
discrepancy between \masssig\ and \fluxsig\ was likely related to the dynamically extreme
nature of this snapshot. For measurements that neglected dust attenuation, the mean
fractional offsets were respectively $-0.06$ and $-0.07$ for \gaussfs\ and~\hermitefs.

The mean fractional offsets presented above were computed using a very broad variety of snapshots.
If we limit the sample to dynamically passive (i.e., non-merging) systems by removing snapshots 1--4 from the analysis,
the fractional offsets of \gaussfs\ become $0.11$ and $-0.15$ for dusty and dustless
systems, respectively. The corresponding fractional offsets of \hermitefs\ become $0.05$
and~$-0.16$.

Many of the individual offsets exceeded the threshold set by the intrinsic measurement
discrepancy. The mean offsets typically exceeded the threshold as well. Two general
trends were observed:

\begin{itemize}
    \item Dust attenuation often caused \fluxsig\ to be elevated with respect to \masssig.
    \item When dust attenuation was neglected, \fluxsig\ tended to be smaller than \masssig.
\end{itemize}

\noindent
Note that many more snapshots from a larger variety of simulations would be needed in
order to compute robust values for these offsets. Also note that these offsets were
observed in a set of galaxies with ongoing star formation. Consequently, the trends in the
offset only strictly apply to systems with ongoing or recent star formation. The trends
are discussed further in the following sections.

The reader may have noticed that the positive offset due to the presence of dust was
smaller, on average, than the negative offset present when dust was ignored.
This occurred even though the flux-weighted measurement technique had an intrinsic
positive offset. We do not consider this to be a robustly-determined trend because the
magnitude of the offset due to dust could easily depend upon the amount of gas and dust
present in the simulation snapshot. The magnitude may also depend upon the type of dust
present in the system or upon the resolution of the underlying simulation. Furthermore,
the relative fraction of young stars likely also affects the magnitude of the offset.
Using our limited sample, we can only determine very general trends, such as
the \textit{direction} of the offsets, with confidence.

\subsection{Dust Attenuation \label{section:attenuation}}

The analysis in Section~\ref{section:ms_vs_fs} revealed that the presence of dust
generally increased \fluxsig, relative to \masssig. From the results of SC2, we also know
that younger stars in merger simulations exhibited lower \masssig\ than older stars,
\textit{on average}. Based on this evidence, it seems likely that dust
\textit{preferentially} obscured the dynamically cooler young stars. This effectively
removed the dynamically cooler stars from the measurement of \fluxsig\ and thus increased
the measured value. This conclusion is consistent with the well-known fact that stars
tend to be born in dusty environments and is further supported by the snapshot renderings
in the \hyperref[section:appendix]{Appendix}, which show that the dust and young stars
tended to be concentrated in the same regions of the system.

In principle, dust can also increase \fluxsig\ measurements by scattering the light of
high velocity stars into the line of sight of the camera, as described by
\citet{baes2002}. This effect is likely weak in our observations because all of
our measurement slits are relatively small and centred upon galactic nuclei, while the
effect of scattering is most important in the outer regions of galaxies.

There were a few exceptions to the general trend, discussed above; some measurements of
\fluxsig\ that included dust attenuation were negatively offset with respect to \masssig.
These exceptions likely occurred in cases for which the attenuation due to dust was weak
or the dust did not attenuate the light of young stars preferentially. The attenuation of
young stars may have been weak for three reasons:

\begin{enumerate}
    \item The total amount of dust in the system was small.
    \item The dust and young stars were distributed in a flattened sheet (or disk) and
the observation was made along a line of sight nearly perpendicular to the sheet.
    \item  The dust became decoupled from the young stars, due to a collision, heating, or
winds.
\end{enumerate}

\noindent
In the first case, the total degree of dust attenuation would be low. In the second and
third cases, the total degree of attenuation could be significant, but the light of
young stars would not be \textit{preferentially} attenuated.

Since we obtained dust-attenuated and dust-free images for each camera direction, we were
able to determine the level of attenuation in each slit. We defined attenuation as,

\begin{equation}
\mathcal{A} = \log\left(\frac{F_{\rm tot}}{F_{\rm dust}}\right), \label{eqn:attenuation}
\end{equation}

\begin{figure}
\centering
    \narrowfigure{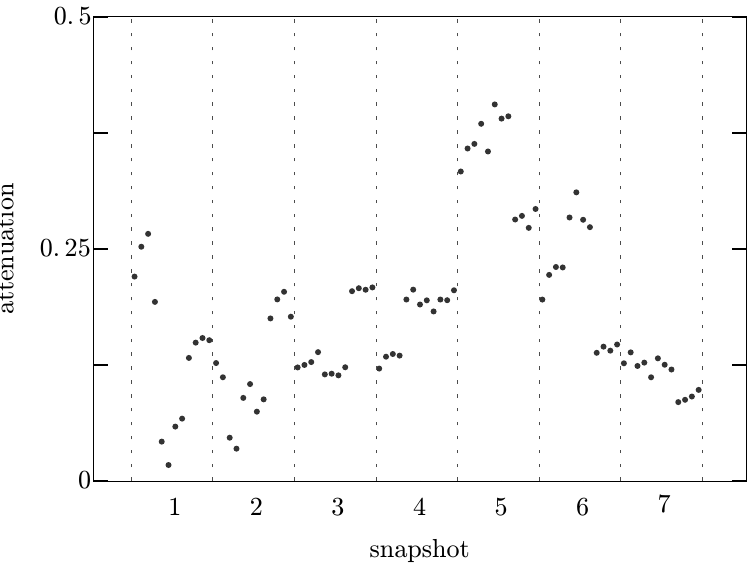}{0.95}
\caption[Dust attenuation of each slit]{The attenuation, $\mathcal{A}$, for each measurement slit.}
\label{figure:attenuation}
\end{figure}
\begin{figure}
\centering
    \narrowfigure{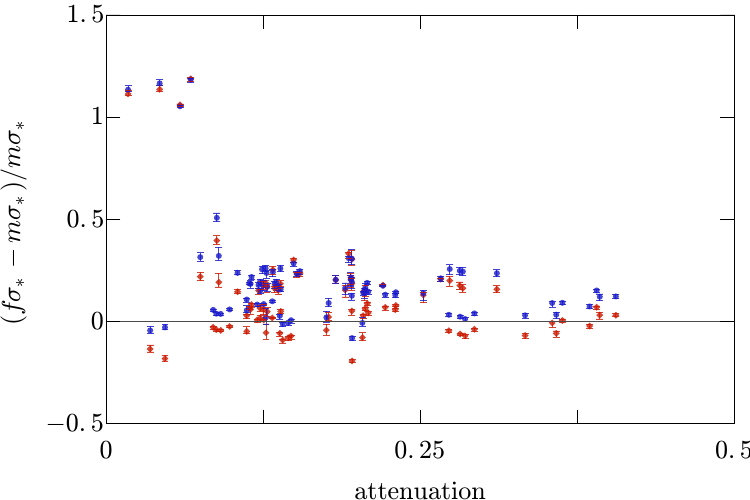}{1.0}
\caption[Flux-weighted offset versus attenuation]{The fractional offset of \fluxsig\ with respect to \masssig\ versus
attenuation, $\mathcal{A}$. It is clear that higher attenuation does not correlate with a larger offset.}
\label{figure:offset-vs-attenuation}
\end{figure}


\noindent
where $F_{\rm tot}$ is the total flux passing through the measurement slit when dust was
ignored and $F_{\rm dust}$ is the flux received when dust was included in the \sunrise\
analysis. The value of $\mathcal{A}$ for all measurement slits is presented in
Figure~\ref{figure:attenuation}. By comparing this figure with Figure~\ref{figure6}, one
can easily see that the total attenuation was not unusually low in situations for which
the dust-attenuated measurement of \fluxsig\ was negatively offset, relative to \masssig.
Furthermore, when the fractional offset, $(f\!\sigmath-m\sigmath)/m\sigmath$, is plotted
as a function of attenuation, as in Figure~\ref{figure:offset-vs-attenuation}, no clear
relationship is evident; total attenuation did not correlate with the fractional offset of
\fluxsig\ from \masssig. We conclude that the distribution of dust with respect to the
distribution of young stars was more important than the total amount of dust present in
the simulations. In other words, the effect of dust on measurements of \fluxsig\ can not
be determined by the amount of reddening present.

\subsection[The Velocity Dispersion of Young Stars]{The \masssig\ of Young Stars \label{section:young-stars}}

In this section, we investigate the exceptions to the second trend observed in
Section~\ref{section:ms_vs_fs}. Specifically, these were cases in which the dust-free
\fluxsig\ was positively offset with respect to \masssig. In principle, the lower
$\Upsilon$ (\ie\ higher luminosity per unit mass) of the young stellar populations should
cause any flux-weighted quantity to shift toward the characteristic value exhibited by the
young populations (when dust attenuation is neglected). It is then reasonable to assume
that the offsets observed in the dust-free \fluxsig\ measurements were due to offsets
present in the population of young stars. We measured the mass-weighted velocity
dispersion using only the relatively young stellar populations in order to determine
whether the observed offsets were due to the dynamics of the young stars.

\begin{figure}
\centering
    \narrowfigure{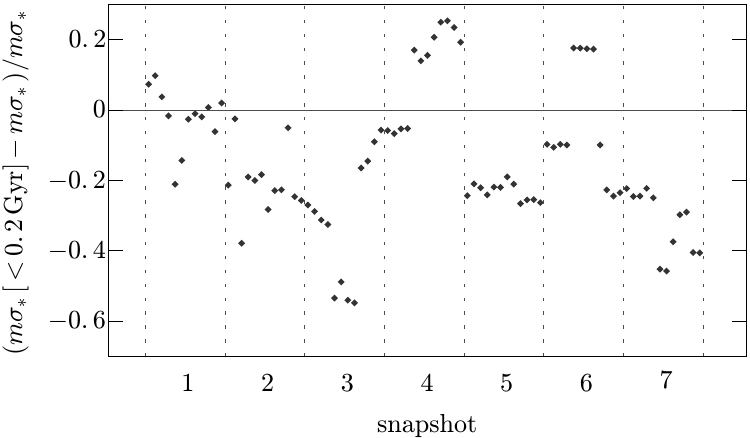}{1.0}
\caption[Fractional offset of \masssig\ in young stars]{The fractional offset of \masssig\ for
stars younger than 200~Myr, relative to the \masssig\ of \textit{all} stars. The offset tends to be 
negative, indicating that the velocity dispersion of younger stars tends to be lower than that of 
older stars.}
\label{figure:young-stars}
\end{figure}

In Figure~\ref{figure:young-stars}, we show the fractional offset of \masssig\ for
stars less than 200~Myr old, relative to the global population. A positive offset was
measured in 17 of the 84 individual measurements. Most of the positively-offset dust-free
\fluxsig\ measurements in Figure~\ref{figure6} can be easily explained using the
data presented in Figure~\ref{figure:young-stars}; the measurements of \fluxsig\ were
positively offset because the \sig\ of young stars was elevated. Computing \masssig\ in an
appropriate stellar age bin (\ie\ something other than 0--200~Myr) would
likely explain the remaining positive offsets.

While young stars were dynamically cooler than the global population, \textit{on average},
individual measurements of \masssig\ in the young population were able to exceed the
global value. This occurred in dynamically peculiar situations (\eg\ in snapshots~1 and 4,
which were galactic collisions) and also when young stars were in a rotating disk, viewed
edge-on (as in the case of Camera~2 of snapshot~6).

\section{DISCUSSION}\label{section:discussion}

By analysing snapshots from a simulation of merging disk galaxies, we were able to compare
mass-weighted velocity dispersion measurements (\masssig) with flux-weighted velocity
dispersion measurements (\fluxsig). The measurements of \fluxsig\ were performed by
generating synthetic spectra with the radiative transfer code, \sunrise. The spectra were
then analysed using the pPXF code of \citet{cappellari2012}---a code which is commonly
used to measure velocity dispersions from real (i.e., observationally-obtained) spectra.
Two spectra were obtained for each observation slit---one that included the effect of dust
attenuation and one that ignored the presence of dust. All of our simulated spectra
included effects of stellar evolution. Our primary findings were:

\begin{enumerate}
    \item Dust preferentially obscured the light of young stars because young stars were
    often found in dusty environments. Therefore, dust partially removed the dynamically
    cool young stars from the measurement of \fluxsig, while stars of all ages
    contributed equally to the \masssig\ measurement. This caused measurements of
    \fluxsig\ to be elevated with respect to \masssig\ in most cases.

    \item When dust was ignored, measurements of \fluxsig\ tended to fall below their
    \masssig\ counterparts. This was due to the higher luminosity-to-mass ratio (i.e.,
    smaller $\Upsilon$) of the young stellar populations, which weighted \fluxsig\ toward
    the velocity dispersion of the dynamically cooler young stars.
\end{enumerate}

\noindent
In exceptional cases, the dust-free measurements of \fluxsig\ exceeded \masssig\ or the
dust-attenuated measurements of \fluxsig\ fell below \masssig. The former type of
exception occurred when the dynamics of the young stars were peculiar or when the system
was observed along a fortuitous line of sight that caused the velocity dispersion of the
young population to appear elevated. The latter type of exception occurred when the dust
did not obscure the young stellar population more significantly than it obscured the older
population. We intentionally chose snapshots and viewing directions that we knew
would increase the likelihood of finding such exceptional cases.

We also found that the total degree of attenuation due to dust was not a good predictor of
the offset between \fluxsig\ and \masssig. Observations with significant dust attenuation
often exhibited a smaller offset due to dust than observations that suffered less
attenuation. In other words, the distribution of the dust was more important than the
total amount of dust present. Results of previous research indicated that
diffusely-distributed dust can have the opposite effect; it can lower the value of
\fluxsig\ relative to \masssig. In SC1, we found that the presence of a large attenuating
slab resulted in measurements of \fluxsig\ that were lower than \masssig. This happened
because the stars of highest dispersion (in the centre of the galaxy) were more strongly
attenuated. \citet{baes2000} found similar results when analytically studying diffusely
distributed dust in elliptical galaxies. Thus, dust may have opposing effects on
measurements of \fluxsig, depending on how it is distributed. We were unable to test this
with the simulated observations described in this paper because the dust was not purely
diffusely distributed.

When comparing spectral modelling methods, we found that using a Gauss-Hermite series to
model the LOSVD yielded lower values of \fluxsig, on average, compared with a pure
Gaussian LOSVD when dust was included in the measurement. There was no systematic offset
between the Gauss and Gauss-Hermite methods when dust was ignored.

Based on our findings, we can provide some general advice regarding comparisons between
observationally-obtained measurements of \sig\ (which is really \fluxsig) and
measurements of \masssig\ in the galaxy simulation literature:

\begin{itemize}
 \item In systems that appear to be passively evolving (i.e., systems that clearly
contain only one nucleus and little or no tidal debris), most measurements of \fluxsig\
are likely to fall within 20\% of the \masssig\ measurement, regardless of the amount of
dust present in the system. Even in extreme cases, \fluxsig\ should not differ from
\masssig\ by more than 30\% in such systems.

\item Measurements of \fluxsig\ in passively evolving gas-rich systems with recent or
ongoing star formation are most likely slightly positively offset with respect to
\masssig.

\item If a large number of young stars (younger than \mbox{$\sim200$}~Myr) in a passively
evolving galaxy are not significantly obscured by dust, a measurement of \fluxsig\ is
likely to be slightly negatively offset with respect to \masssig.

\item In systems that are actively merging (i.e., systems with two or more visible
nuclei, more than one large disk structure, or an otherwise disturbed appearance),
\fluxsig\ can differ from \masssig\ by more than 100\% in extreme cases. 
\end{itemize}

\noindent
Finally, we note some important caveats to our findings:

\begin{itemize}

\item The numerical simulations that were used to create the galaxy snapshots were not
perfect. Notably, the resolution limit was 25~pc, which means that the structure of the
ISM was not resolved on scales of $\lesssim 25$~pc. This also means that the particles in
the simulation and the subsequent \sunrise\ radiative transfer scheme represented entire
stellar populations, with no sub-structure.

\item We simulated a small region of the spectrum (5040~\AA\ to 5430~\AA). The results may
differ somewhat in different regions of the spectrum. On the other hand, this region is
commonly used in real observations of velocity dispersion, so it is a useful choice.

\item Changing the dust grain model, dust-to-metal ratio, or total gas dust in the systems
would have likely effected the details of the results. The magnitudes of the offsets would
likely be different if any of these parameters were changed, however the direction of each
offset is likely insensitive to to these variables.

\item Our observational sample was quite small and it was not randomly chosen. It included
some snapshots with highly unusual dynamics and specially-selected viewing directions.
The extreme nature of the sample allowed us to identify exceptional cases which might not
have been observed if the snapshots and viewing directions were chosen at random. However,
one cannot compute robust statistics from this sample.

\end{itemize}

\section*{ACKNOWLEDGEMENTS}

We thank Gillian Wilson and Desika Narayanan for their invaluable assistance. This work
used the Extreme Science and Engineering Discovery Environment (XSEDE), which is supported
by National Science Foundation grant number OCI-1053575. Financial support for this work
was provided by NASA through a grant from the Space Telescope Science Institute (Program
numbers AR-12626 and GO-11557), which is operated by the Association of Universities for
Research in Astronomy, Incorporated, under NASA contract NAS5-26555.

\bibliographystyle{mnras}
\bibliography{references}

\appendix


\section{APPENDIX: DESCRIPTIONS OF OBSERVATIONS} 
\label{section:appendix}

In this appendix, we describe each of the 21 individual observations (snapshot
and camera position).

\subsubsection*{Snapshot 1}

\begin{figure*}
    \widefigure{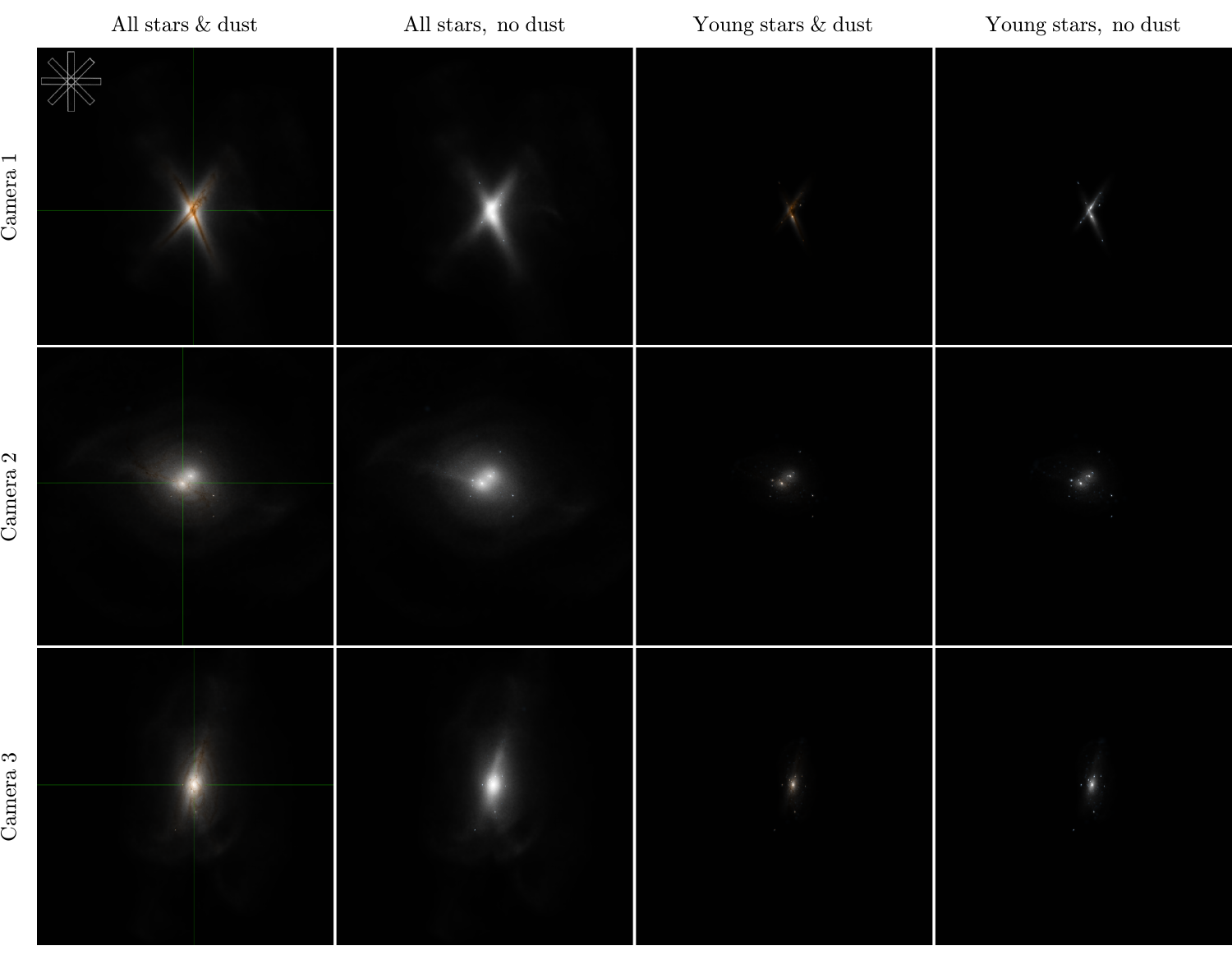}{0.95}
\caption[Visualizations of Snapshot 1]{\label{fig:snap1}\GSnap-generated renderings of
Snapshot~1 ($t=2.060$~Gyr), viewed along the three camera directions that were examined.
Each image represents a $100\times100$~kpc region. Green cross-hairs indicate the position
of the centre of the slits. The size and orientation of the slits is shown in the
upper-left corner, for reference. The first column shows the system with all stars and
dust dust attenuation included. The second column shows the system without dust
attenuation. The third column shows only the stars that formed during the simulation,
attenuated by dust. The fourth column shows the image in the third column in the absence
of dust attenuation. The brightness scaling in all images is identical.}
\end{figure*}

\noindent
This snapshot was recorded at $t = 2.06$~Gyr---immediately before the climax of the second pass (the 
first pass occurred at 0.49~Gyr). We knew from the analysis performed in SC2 that the mass-weighted 
velocity dispersion reached its highest value during this stage of the merger. Consequently, this 
was a dynamically extreme system. The system contained two disks which were in the process of 
passing through one another. See Figure~\ref{fig:snap1} for renderings of the system from the three 
camera positions and refer to the upper two panels of Figure~\ref{summaryplot} to see how this 
snapshot fits into the overall evolution of the merger.

Camera~1 was placed along a line of sight containing maximal dust attenuation. This line
of sight fell approximately $27^\circ$ away from the collision axis. Camera~2 was placed
nearly perpendicular to the collision axis along a line of sight with minimal dust
extinction. Camera~3 observed an intermediate configuration with a moderate amount of
attenuation due to dust and a viewing angle $33^\circ$ from the collision axis.

\subsubsection*{Snapshot 2}

\noindent
This snapshot was recorded at $t = 2.089$~Gyr. Dynamically, the system was still somewhat
excited, having recently undergone a major collision. The system consisted of two clearly
distinguishable, disturbed ellipsoids with extended disk debris.

\begin{figure*}
\centering
    \widefigure{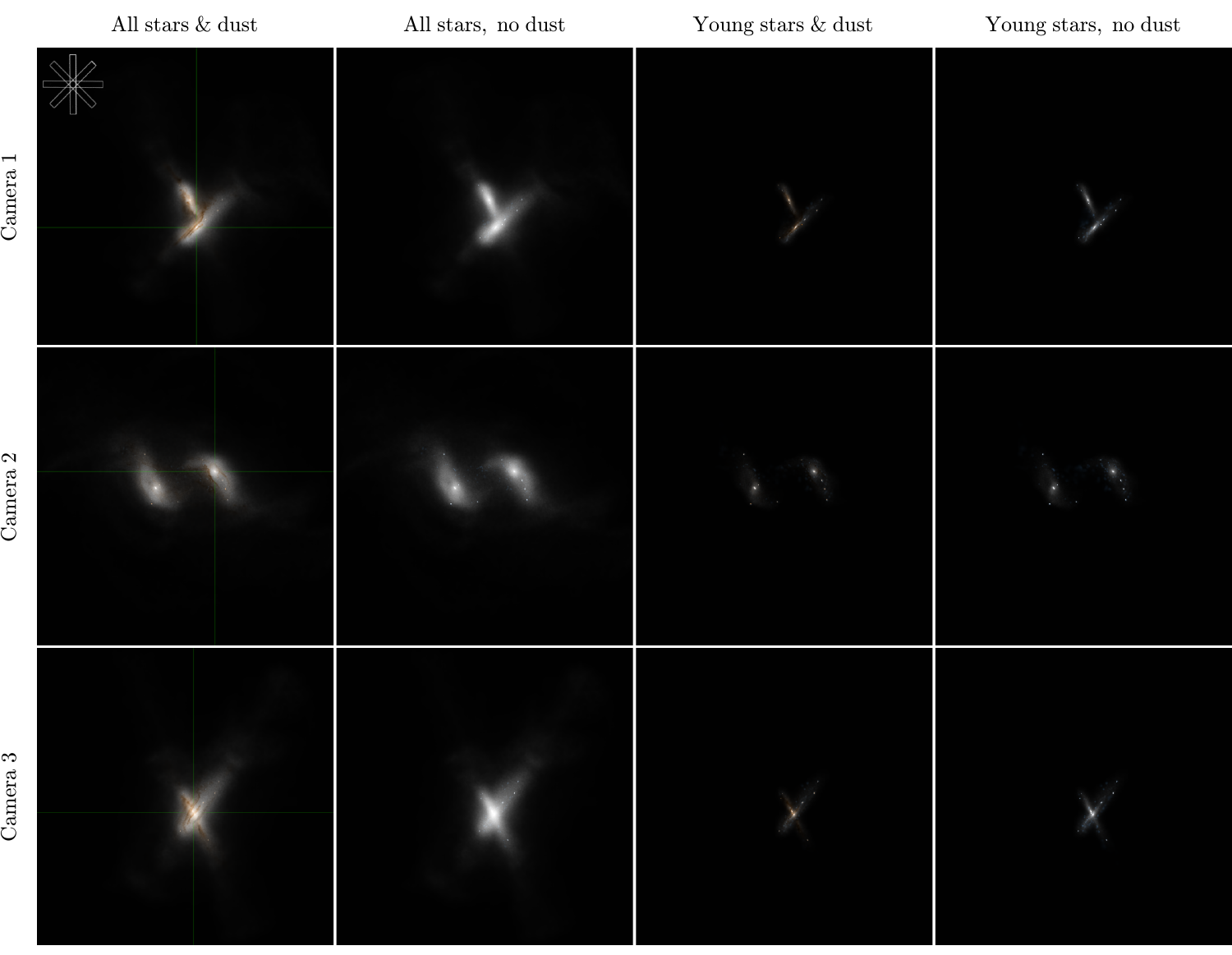}{0.95}

\caption[Visualizations of Snapshot 2]{\label{snap2}The same as Figure~\ref{fig:snap1}, but for Snapshot~2
($t=2.089$~Gyr).}
\end{figure*}

Camera~1 was placed along a line of sight with a moderate degree of dust attenuation,
$24^\circ$ from the collision axis. Camera~2 was placed along a line of sight
perpendicular to the collision axis with moderate attenuation. Camera~3 was placed
approximately $5^\circ$ from the collision axis. Attenuation was somewhat more significant
in the Camera~3 observation than in the other two. A significant number of stars from both
spheroids appeared in all slits of Cameras~1 and 3, whereas the slits used in the Camera~2
measurements primarily included stars belonging to only one of the ellipsoids. See
Figure~\ref{snap2} for renderings.

\subsubsection*{Snapshot 3}

\noindent
This snapshot was recorded at $t = 2.181$~Gyr, which is approximately midway between the
second and third passes of the merger process. The system consisted of two disturbed
ellipsoidal galaxies. Both ellipsoids had nearly reached a dynamically stable state at
this point. Enhanced star-formation, which was triggered during the second pass, was
recently quenched by periods of quasar activity.

\begin{figure*}
\centering
    \widefigure{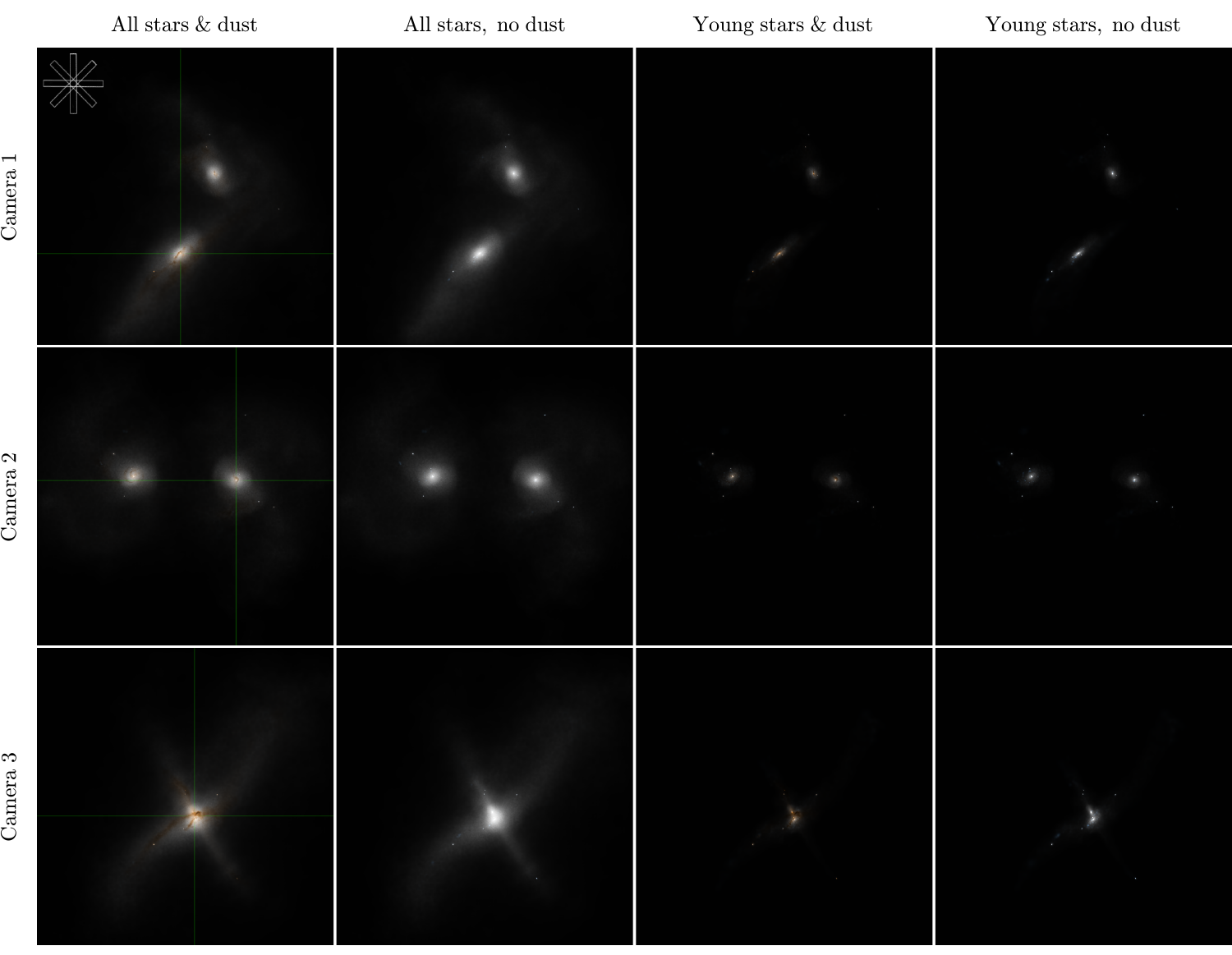}{0.95}
\caption[Visualizations of Snapshot 3]{The same as Figure~\ref{fig:snap1}, but for Snapshot~3
($t=2.181$~Gyr).\label{snap3}}
\end{figure*}

Cameras~1 and 2 were placed along lines of sight with moderate dust attenuation. Both
sub-systems were clearly
distinguishable from these camera positions. Camera~3 was placed along the line of sight
connecting the two systems.
Attenuation was more significant along this camera direction than along the first two
camera directions. See
Figure~\ref{snap3} for renderings.

\newpage
\subsubsection*{Snapshot 4}

\noindent
This snapshot was recorded at $t = 2.299$~Gyr---the climax of the third pass. The two
progenitor systems were fully superimposed; only one nucleus could be identified. Like
Snapshot~1, this snapshot represented a highly unusual system, partially due to its
elevated velocity dispersion, but also because the system hosted a weak nuclear starburst
with a star formation rate of $\sim5~\rm M_\odot~yr^{-1}$.

\begin{figure*}
\centering
    \widefigure{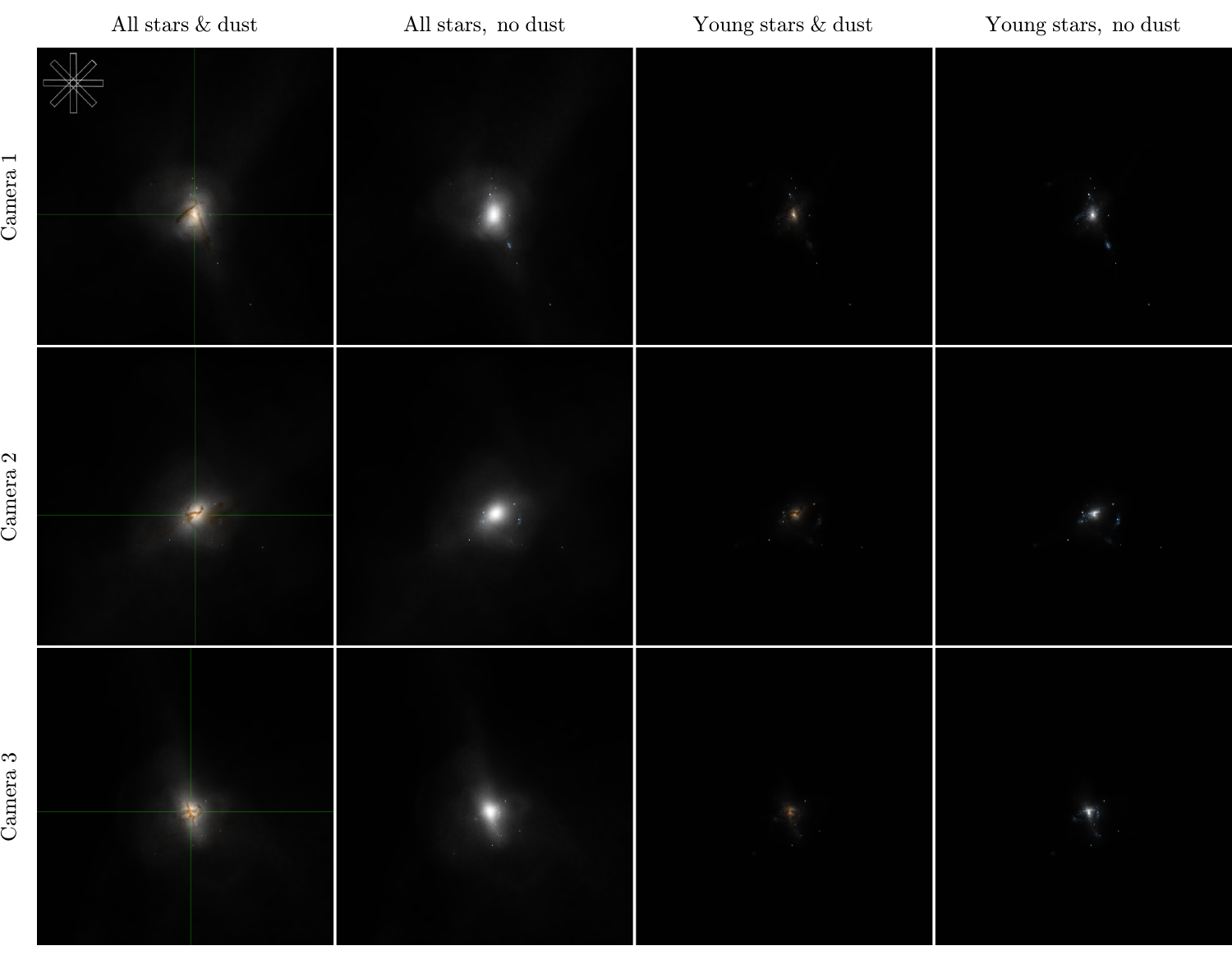}{0.95}
\caption[Visualizations of Snapshot 4]{The same as Figure~\ref{fig:snap1}, but for Snapshot~4 ($t=2.299$~Gyr).
\label{snap4}}
\end{figure*}

Cameras~1, 2, and 3 were positioned $29^\circ$, $38^\circ$, and $40^\circ$ from the
collision axis. Dust attenuation was moderate in all three cases. See Figure~\ref{snap4}
for renderings.

\newpage
\subsubsection*{Snapshot 5}

\noindent
This snapshot was recorded at $t = 2.401$~Gyr---the moment of nuclear coalescence. The
majority of the merger's dynamical evolution was complete at this point. A nuclear
starburst with a star formation rate of $\sim13~\rm M_\odot~yr^{-1}$ was present. Major
quasar activity began $\lesssim5$~Myr after this snapshot was recorded, causing a sharp
decrease in the star formation rate. Therefore, this system was likely similar to some
quasar host galaxies.

\begin{figure*}
\centering
    \widefigure{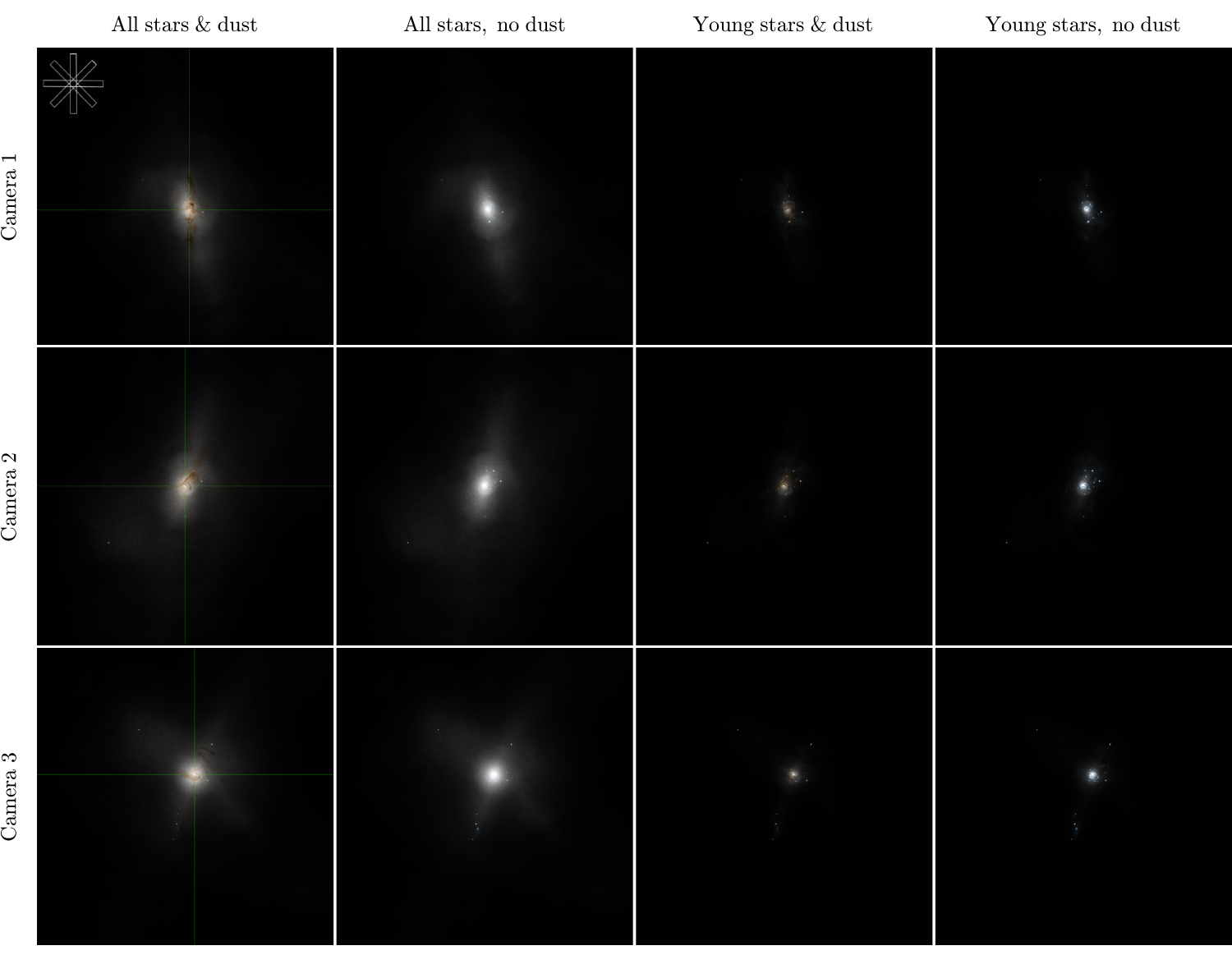}{0.95}
\caption[Visualizations of Snapshot 5]{The same as Figure~\ref{fig:snap1}, but for Snapshot~5
($t=2.401$~Gyr).\label{snap5}}
\end{figure*}

Since the system was nearly isotropic dynamically as well as in terms of its dust
distribution, the camera positions were chosen essentially at random. We used a large
angular separation between viewing directions in order to prevent us from sampling the
same regions of the simulation more than once. See Figure~\ref{snap5} for renderings of
Snapshot~5.

\newpage
\subsubsection*{Snapshot 6}

\noindent
This snapshot was recorded at $t = 2.690$~Gyr. The system exhibited shells and tidal debris in its 
outer regions. It also contained two non-coaxial, concentric nuclear disks with diameters of 0.3~kpc 
and 1.3~kpc. The inner disk was inclined $67^\circ$ with respect to the outer disk.

\begin{figure*}
\centering
    \widefigure{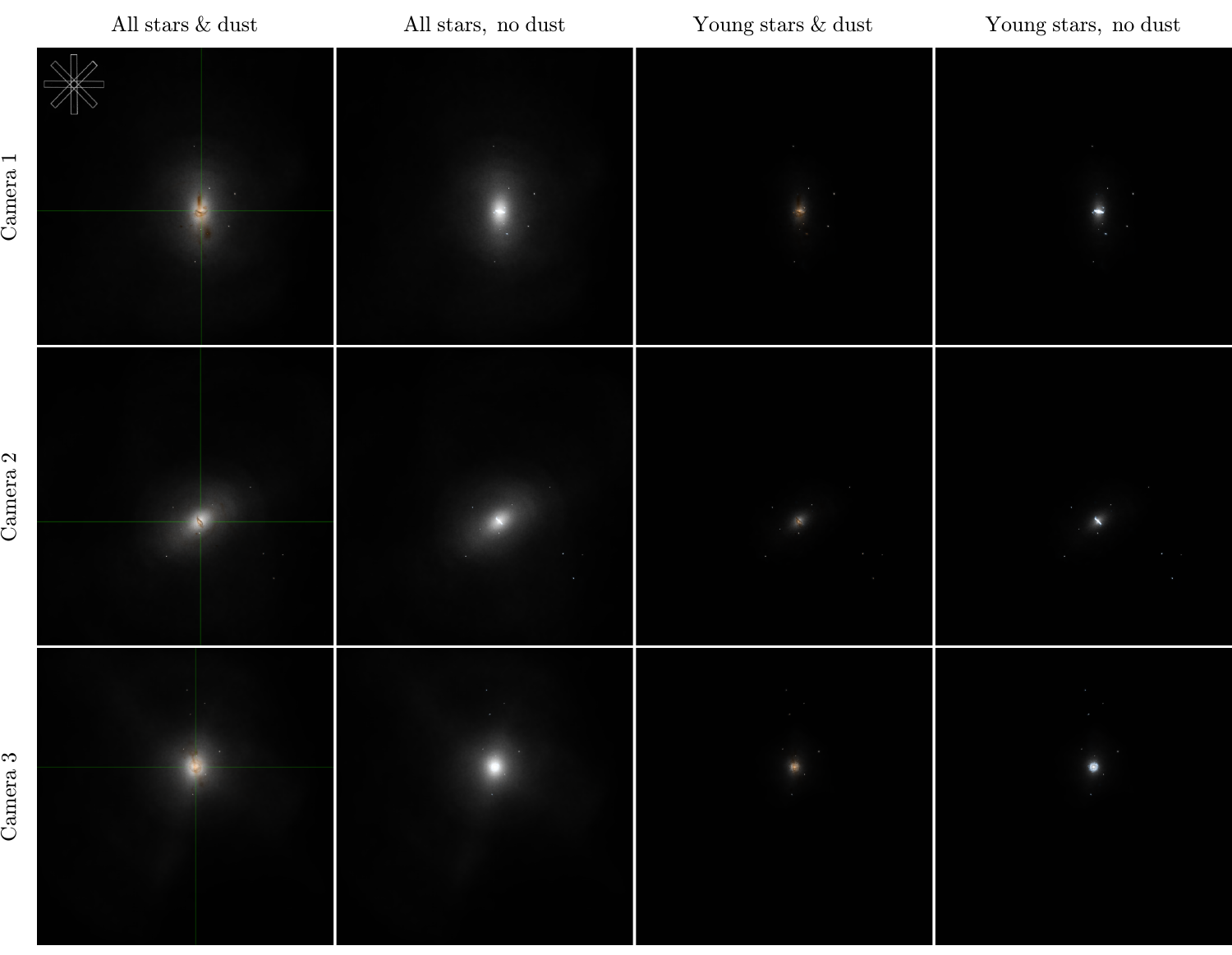}{0.95}
\caption[Visualizations of Snapshot 6]{The same as Figure~\ref{fig:snap1}, but for Snapshot~6 ($t=2.690$~Gyr).
\label{snap6}}
\end{figure*}

Camera~1 was placed along a line of sight inclined $\sim 40^\circ$ with respect to the direction 
defined by the intersection of the planes containing the two disks. This inclination angle was 
measured along a third plane that symmetrically bisected the system. In other words, the two disks 
were viewed from equal inclination angles. Camera~2 was placed such that both disks were viewed 
edge-on. Camera~3 was placed such that the larger disk was viewed edge-on. See Figure~\ref{snap6} 
for renderings.

\newpage
\subsubsection*{Snapshot 7}

\begin{figure*}
\centering
    \widefigure{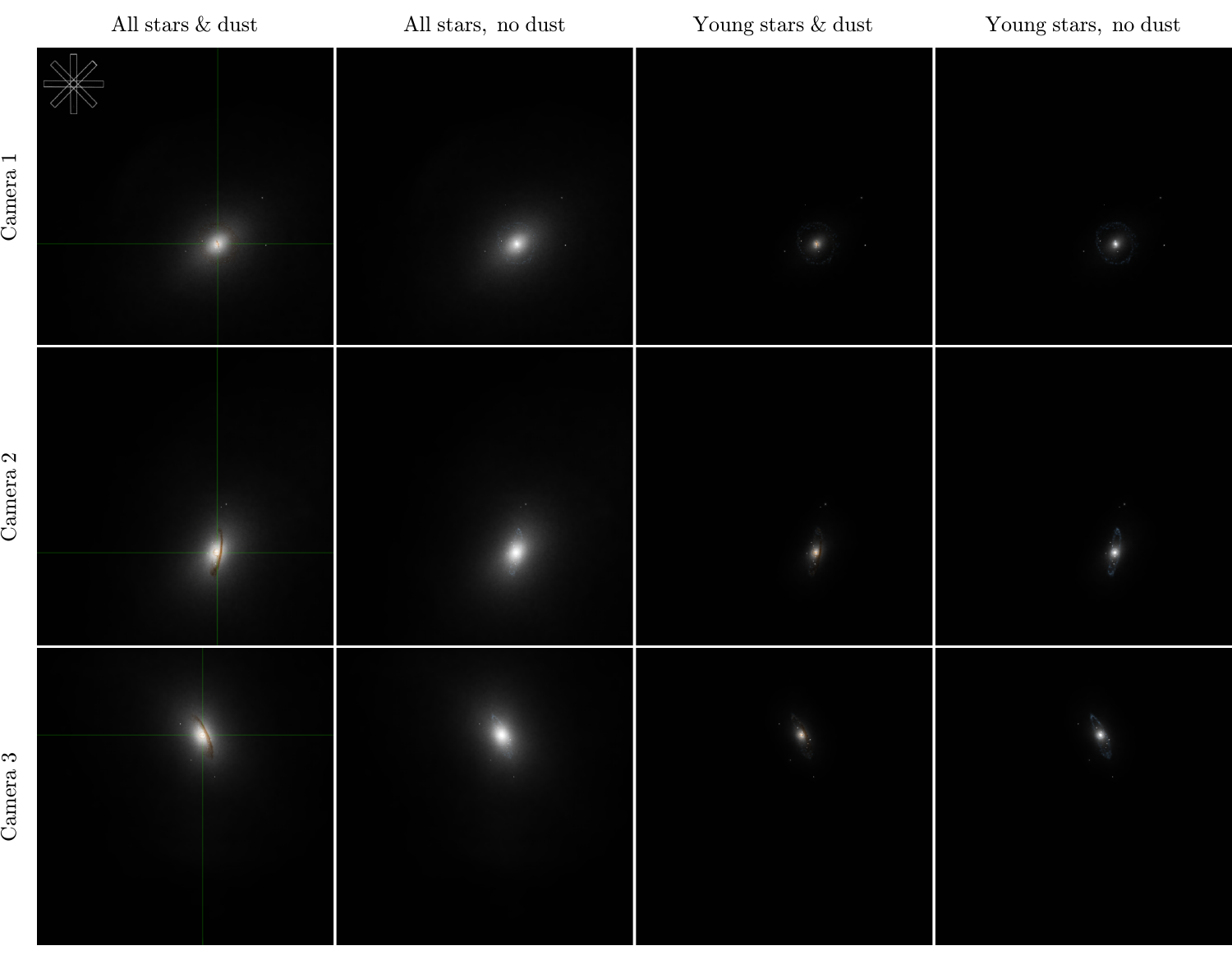}{0.95}
\caption[Visualizations of Snapshot 7]{The same as Figure~\ref{fig:snap1}, but for Snapshot~7
($t=3.620$~Gyr).\label{snap7}}
\end{figure*}

\noindent
This snapshot was recorded at $t = 3.62$~Gyr---the end of the simulation. The system
consisted of an elliptical galaxy containing three small nuclear disks with diameters of
9.0~kpc, 1.5~kpc, and 0.3~kpc. It may be more appropriate to refer to the largest disk,
as a ``ring,'' since it clearly exhibited outer and inner edges. See
Figure~\ref{snap7} for
renderings.

Camera~1 viewed the large ring face-on; the intermediate disk was inclined $\sim20^\circ$ with 
respect to this line of sight. Camera~2 viewed the small inner disk edge-on. Camera~3 viewed the 
intermediate disk face-on. Cameras~2 and 3 respectively fell along lines of sight inclined 
$\sim20^\circ$ and $\sim23^\circ$ with respect to the edge of the large ring.

\bsp	
\label{lastpage}
\end{document}